\documentclass[12pt]{article}

\IfFileExists{srcltx.sty}{\usepackage[active]{srcltx}}

\usepackage{amssymb,amsmath,bm,array}%
\usepackage[dvips]{graphicx}%

\newcount\hour \newcount\minute \hour=\time \divide \hour by 60 \minute=\time
\usepackage{xspace} %
\usepackage{paralist} %
\usepackage[a4paper,hscale=.8,vscale=.75,centering]{geometry} %

\usepackage[colorlinks=true]{hyperref}

\let\oldbox=\Box %
\renewcommand{\Box}{\mathop{\oldbox}}

\let\oldint=\int %
\renewcommand{\int}{\mathop{\hskip -.25ex\oldint \hskip -.4ex}}

\usepackage[colorlinks=true]{hyperref}

\newcommand{\EM}{\textsc{em}} 
\newcommand{\cs}{\textsc{cs}} 

\newcommand{\p}[1]{\partial_{#1}} %
\renewcommand{\u}[1]{\ensuremath{\mathrm{U}(#1)}\xspace} 
\renewcommand{\v}[1]{\ensuremath{v_{#1}}} %
\newcommand{\vb}{\ensuremath{v_{B}}} %
\newcommand{\fg}{\ensuremath{F_{A}}} %
\newcommand{\fb}{\ensuremath{F_{B}}} %
\newcommand{\q}[1]{\ensuremath{q_{#1}}}
\renewcommand{\k}{\ensuremath{\bm{\kappa}}\xspace} %
\newcommand{\kz}{\ensuremath{k_z}} %
\newcommand{\m}[1]{\ensuremath{m_{#1}}} %
\newcommand{\mb}{\ensuremath{m_{B}}}
\newcommand{\ev}{~\text{eV}} %
\newcommand{\gev}{~\text{GeV}} %
\newcommand{\alp}{\textsc{alp}} 
\newcommand{\CL}{\mathcal{L}} 

\begin{document}
\author{\large \sf I.~Antoniadis~$^{a,}$\thanks{On leave from CPHT
(UMR CNRS 7644) Ecole Polytechnique, F-91128 Palaiseau}~,
~~A.~ Boyarsky~$^{a,b,}$\thanks{On leave of absence from Bogolyubov
Institute of Theoretical Physics, Kyiv, Ukraine}~,
~~Oleg Ruchayskiy~$^{c}$\\[0.5cm]
$^a$\normalsize\emph{Department of Physics, CERN - Theory Division, 1211 Geneva 23,
Switzerland}\\
$^b$\normalsize\emph{Ecole Polytechnique F\'ed\'erale de
    Lausanne, Institute of Theoretical Physics}\\
\normalsize\emph{FSB/ITP/LPPC, BSP 720,
    CH-1015, Lausanne, Switzerland}\\
$^c$\normalsize\emph{Institut des Hautes \'Etudes Scientifiques, Bures-sur-Yvette, F-91440, France}
} %
\title{\begin{flushright}
\small CERN-PH-TH/2006-119
\end{flushright}Axion alternatives} %
\maketitle

\begin{abstract}
  If recent results of the PVLAS collaboration proved to be correct, some
  alternative to the traditional axion models are needed. We present one of
  the simplest possible modifications of axion paradigm, which explains the
  results of PVLAS experiment, while avoiding all the astrophysical and
  cosmological restrictions. We also mention other possible models that
  possess similar effects.
\end{abstract}

\section{Introduction}
\label{sec:introduction}

\begin{figure}[t]
  \centering
  \includegraphics[width=.5\textwidth]{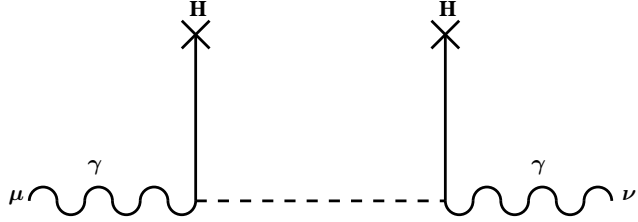} %
  \caption{Diagram for effective mass of a photon in the external magnetic
    field $H$. The dashed line corresponds to an exchange of the hypothetical new particle.} %
  \label{fig:propagator}
\end{figure}

It has been understood long ago, that the presence of light pseudoscalar
particles, with coupling to the electromagnetic field of the form
\begin{equation}
  \label{eq:1}
  \CL_{\textsc{em}+\phi} = -\frac14 F_{\mu\nu}^2 + \frac12(\p\mu\phi)^2 -
  \frac{m^2_\phi}2   \phi^2 - \frac1{4f_\phi}\phi\, \epsilon^{\mu\nu\lambda\rho}F_{\mu\nu} F_{\lambda\rho}\;,
\end{equation}
leads to a number of signature
effects~\cite{Iacopini:79,Sikivie:83,Anselm:85,Gasperini:87,Maiani:86,Raffelt:88}.
In particular, in external magnetic field, vacuum becomes
birefringent and dichroic. These effects (change of linear
polarization into the elliptic one and rotation of the
polarization plane of linearly polarized light) can be
qualitatively understood as follows. Interaction~(\ref{eq:1})
leads to an effective mass for the component of the photon,
polarized along the direction of the external magnetic field (see
Fig.\ref{fig:propagator}), while perpendicularly polarized
component remains massless. Recently, PVLAS collaboration reported
observation of such an effect of rotation of
polarization~\cite{PVLAS}.

However, the axion interpretation of the PVLAS data comes in contradiction
with other experimental constraints~(see e.g.~\cite{Ringwald:05}). Namely,
PVLAS data can be interpreted in terms of axion-like particle (\emph{ALP})
with Lagrangian~(\ref{eq:1}) if one identifies
\begin{equation}
  \label{eq:2}
  m_\phi = 1-1.5\;\mathrm{meV}; \qquad f_\phi = (2-6)\times 10^5\gev
\end{equation}
These numbers disagree, however, with both astrophysical bounds
(see e.g.~\cite{Raffelt:99})
\begin{equation}
  \label{eq:3}
  f_{a,\,\mathrm{stellar}} \gtrsim 10^{10}\gev
\end{equation}
and the results of CAST collaboration~\cite{CAST}, which tries to detect the
flux of $\phi$ coming from the Sun and its conversion into photons in a
strong magnetic field. Absence of such a signal puts a lower bound on $f_\phi$,
compatible with~(\ref{eq:3}):
\begin{equation}
  \label{eq:4}
  f_\phi > 8.6\times 10^{9}\gev\quad\text{for}\quad m_\phi \lesssim 0.02\ev
\end{equation}
Clearly, there exists a strong contradiction between the results
of PVLAS and limits~(\ref{eq:3})--(\ref{eq:4}), which implies that
the simple axion model~(\ref{eq:1}) should be modified. Attempts
to avoid the astrophysical constraints~(\ref{eq:3})--(\ref{eq:4})
have been discussed in the literature~(see
e.g.~\cite{Masso:05,Jain:05,Masso:06}). We note that models with
ALPs are historically motivated by the CP problem in QCD and the
Peccei-Quinn mechanism of its solution~\cite{strongCP,axion}.
Therefore the attempts to avoid stellar constraints are usually
based on modification of the axion Lagrangian, making the theory
more sophisticated.

In this work we suggest a different way to reconcile the results of PVLAS
experiment~(\ref{eq:2}) with the constraints (\ref{eq:3})--(\ref{eq:4}). We
present another very simple model which provides almost the same properties of
propagation of photons in the magnetic field as models with axion. This model
is not inspired by the strong CP problem and has rather different particle
physics motivation. We show that in such a model there is no contradiction
between PVLAS data interpreted as a mass of photon in the magnetic field and
astrophysical and CAST constraints. At the end we briefly discuss
high-energy motivation of this model, various modifications of the effective
theory coming from different fundamental theories and possibilities to
distinguish experimentally between them. We plan to discuss these issues in
detail elsewhere~\cite{Antoniadis:06}.

The plan of the paper is the following. In the
section~\ref{sec:massive-photon} we describe the effect of an
effective photon mass generation in a magnetic field and present
the model. In the section~\ref{sec:propagator} we calculate the
propagator of of the photon in a magnetic field in our model.
Finally, in the last section~\ref{sec:discussion}, we shortly
discuss our results, possible theoretical origin of our model and
its future developments.

\section{New vector boson}
\label{sec:massive-photon}

\begin{figure}[t]
  \centering \includegraphics[width=.45\textwidth]{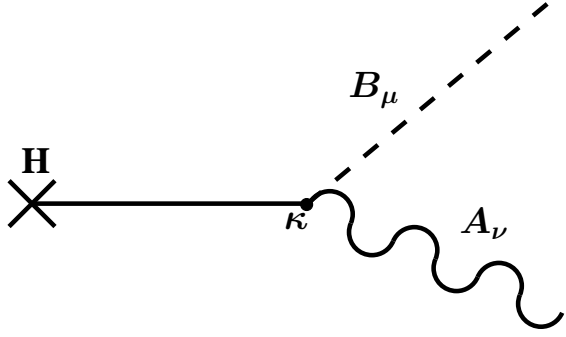}
  \caption{Creation of $B_\mu$ in the external fields}
  \label{fig:vertex}
\end{figure}

The effects of birefringence and dichroism can be qualitatively described as
generation of the effective mass of the photon via the process, shown on
Fig.~\ref{fig:propagator}. Qualitatively, one expects this \emph{magnetic
  mass} to behave as
\begin{equation}
  \label{eq:5}
  m_{\gamma,H}  \sim \frac{\k H}{\m\alp}
\end{equation}
where $\m\alp$ is the mass of intermediate particle (whose propagator is
denoted by dashed line on Fig.~\ref{fig:propagator}) and $\k$ is a
dimensionless coupling constant in the vertex defining the
interaction of photon with this intermediate particle in a constant magnetic
field.\footnote{Indeed, the diagram~\ref{fig:propagator} is equal to $\k^2
  D(p^2)$, where $D(p^2)$ is the propagator of the intermediate particle and
  $p$ is the 4-momentum of incoming photon. As the effective mass is very small,
  $p^2 \ll m^2_\textsc{alp}$, we get~(\ref{eq:5}). We will present the exact
  calculation of this diagram and its corresponding contribution to the effective
  photon mass in the next Section.}
On the other hand, both constraints~(\ref{eq:3}) and~(\ref{eq:4}) stem from
production (annihilation) of real ALPs, i.e. from processes of the type shown in
Fig.~\ref{fig:vertex}.  By choosing $\k$ to be very small one can
arbitrarily suppress them and still have the effect~(\ref{eq:5}) preserved,
provided that $\m\alp$ also goes to zero with $\k\to 0$.

One of the simplest theories with this property is the theory of a gauge boson
$B$ acquiring mass from a Higgs field $\mb = \k \vb$ with $\k$ being its corresponding
charge. Then, one could have an interaction with the photon $A$
via the \emph{Chern-Simons-like} term
\begin{equation}
  \label{eq:6}
  \CL_\cs = \k\,\epsilon^{\mu\nu\lambda\rho} A_\mu B_\nu (\p\lambda
  A_\rho)
\end{equation}
We will see in a moment that the appearance of the same charge $\k$ in front of this interaction
is natural and is dictated by gauge invariance.
Clearly, the term~(\ref{eq:6}) is the close cousin of the axion-like
interaction~(\ref{eq:1}), which is easy to see if
 one substitutes $B_\mu $ by $\p\mu \phi$.
However, the full theories are \emph{not} equivalent as we will show below.

Naively, the theory with the interaction~(\ref{eq:6}) is not gauge invariant,
its A-gauge variation being proportional to $\fg \tilde \fb$, (where by tilde
we denote dual field-strengths). This can, however, be amended and the simplest
theory which possesses the interaction~(\ref{eq:6}) and is gauge invariant is
the following:
\begin{align}
  \label{eq:7}
  \CL & = -\frac 14|\fg|^2 - \frac 14 |\fb|^2 + \frac {\v\gamma^2}2
  (D_\mu\theta_1)^2
  + \frac{\vb^2}2 (D_\mu\theta_2)^2 \\
  & \hphantom{=} + \theta_1 \epsilon^{\mu\nu \lambda\rho} (F_A)_{\mu\nu}
  (F_B)_{\lambda\rho} + \theta_2 \epsilon^{\mu\nu \lambda\rho} (F_A)_{\mu\nu}
  (F_A)_{\lambda\rho} - 2\k\,\epsilon^{\mu\nu\lambda\rho} A_\mu B_\nu
  (F_A)_{\lambda\rho}\notag
\end{align}
where
\begin{equation}
  \label{eq:8}
  D_\mu\theta_1 = \p\mu \theta_1 +\k A_\mu;\quad D_\mu\theta_2 = \p\mu \theta_2 -\k B_\mu
\end{equation}
The theory~(\ref{eq:7}) may seem somewhat \emph{ad hoc}. However, such
interactions appear generically in effective theories of models with
chiral fermions, which acquire mass via Yukawa couplings with the charged Higgs
fields~\cite{DHoker:84a,DHoker:84b}. We will comment on the possible
microscopic origin of the theory~(\ref{eq:7}) in the Discussion section, the detailed
analysis will appear elsewhere~\cite{Antoniadis:06}.  The fields $\theta_1$
and $\theta_2$ are the phases of the Higgs fields, which provide masses
to both gauge fields.
The coefficients in Eq.~(\ref{eq:7}) are fixed by requirement of
gauge invariance with respect to the two \u1 groups: $\u1_\EM\times \u1_B$:
\begin{align}
  \label{eq:9}
  &\u1_\EM: & \delta_\lambda \theta_1 &=- \k \lambda; & \delta_\lambda A_\mu=
  \p\mu\lambda\\
  &\u1_B: &\delta_\beta \theta_2 &= \k \beta; &\delta_\lambda B_\mu = \p\mu\beta
\end{align}
where only non-trivial gauge-variations are shown for each group. The same
theory, considered in the unitary gauge for both fields, reads:
\begin{equation}
\label{eq:10}
  \CL = -\frac 14|\fg|^2 - \frac 14 |\fb|^2 + \frac {\m\gamma^2}2 A_\mu^2 +
  \frac{\mb^2}2B_\mu^2 -2 \k\epsilon^{\mu\nu\lambda\rho} A_\mu B_\nu (
  \fg)_{\lambda\rho}
\end{equation}
where
\begin{equation}
  \label{eq:11}
  \m\gamma = \k \v\gamma;\qquad \mb = \k\vb
\end{equation}

The theory~(\ref{eq:10}) will be analyzed in details in the next Section.
However, one can anticipate the results.  Substituting uniform magnetic field
$H$ in place of $F_A$ in the interaction term~(\ref{eq:10}) and diagonalizing
the quadratic part of the Lagrangian, one can see that the mass of the photon
shifts: $\m\gamma^2 \to \m\gamma^2 + H^2/\vb^2$, the dependence of $\k$
cancels out and the effect is not suppressed in the limit of small $\k$. This allows us to choose $\vb
\sim 10^5\gev$ in order to explain the PVLAS data and choose $\k \lll 1$ to suppress
emission of $B$-boson from stars. Actually, by looking at
Eqs.~(\ref{eq:10})--(\ref{eq:11}), it is clear that \k should be very small.
Indeed, according to the Particle Data Group~\cite{PDG} the upper bound on
photon mass is\footnote{Most of the model-independent constrains
  come from direct measurements of deviation of Coulomb law from $r^{-2}$
  dependence~\cite{Williams:71}, quoted in~\cite{PDG}. There exist much
  stronger experimental restrictions, $\m\gamma < 3\times
  10^{-27}$~eV~\cite{Chibisov:76, Lakes:98}, which are however
  model-dependent~\cite{Adelberger:03}.  }
\begin{equation}
  \label{eq:12}
  \m{\gamma,\textsc{pdg}} < 6\times 10^{-17}\ev\;.
\end{equation}
We will choose the constant \k in such a way, as to ensure the
limit~(\ref{eq:12}) as well as the suppression of $B$ production in stars and in the early
Universe.

\section{Propagator of light in a uniform magnetic field}
\label{sec:propagator}

In this Section we compute the propagator of photon in the model of
Section~\ref{sec:massive-photon} and show that it leads indeed to effects of
birefringence and dichroism.

In the theory~(\ref{eq:10}) the \emph{full} propagator of
a photon $G_{\mu\nu}$ can be
determined from the following equation:\footnote{Our conventions are as
  follows: Greek indices $\mu,\nu,\dots = 0,\dots,3$, Latin indices
  $a,b,\dots=\{0,1\}$ and $i,j,\dots = \{2,3\} $. Metric is mostly-negative
  $\eta_{\mu\nu} = (+1,-1,-1,-1)$, $\epsilon^{ab}$ is an antisymmetric tensor
  with $\epsilon^{01} = 1$. We also use indices $0,1,2,3$ and $t,x,y,z$
  interchangeably.}
\begin{equation}
  \label{eq:13}
  G^{-1}_{\mu\nu} = (G^0)^{-1}_{\mu\nu} -
  \Pi_{\mu\nu}(k)
\end{equation}
where the 1PI self-energy of the photon $\Pi_{\mu\nu}$ is determined by the
diagram~\ref{fig:propagator},
and the \emph{tree level} propagators of the $B$-field and the
photon are given by
\begin{equation}
  \label{eq:14}
    D_{\mu\nu}(k) = \frac{\eta_{\mu\nu} - {k_\mu k_\nu}/{\mb^2}}{k^2 - \mb^2 +
    i\epsilon}; \qquad G^0_{\mu\nu} = \frac{\eta_{\mu\nu} - {k_\mu
      k_\nu}/{\m\gamma^2}}{k^2 - \m\gamma^2 +  i\epsilon}\, .
\end{equation}
$ (G^0)^{-1}_{\hphantom{-1}\mu\nu}$ is the inverse propagator~(\ref{eq:14}):
\begin{equation}
  \label{eq:15}
  (G^0)^{-1}_{\mu\nu}  = (k^2
  -\m\gamma^2)\eta_{\mu\nu} -{k_\mu k_\nu}
\end{equation}
We are interested in the propagation of light in the uniform magnetic field
$H=F_{23}$. Then, $\Pi_{\mu\nu}$ is given by
\begin{equation}
  \label{eq:16}
  \Pi^{ab} = \k^2 H^2\epsilon^{ac}\epsilon^{bd}\left(\frac{\eta_{cd} -
      {k_c k_d}/{\mb^2}}{k^2 - \mb^2 + i\epsilon} \right);\qquad \Pi_{ab}
  = \eta_{aa'}\eta_{bb'}\Pi^{a'b'}
\end{equation}
with all other components equal to zero.  One can rewrite $\Pi_{ab}$
as
\begin{equation}
  \label{eq:17}
  \Pi_{ab} = -\frac{\k^2 H^2}{\mb^2(k^2 - \mb^2)}
  \begin{pmatrix}
    k_x^2 + \mb^2 & k_x\omega\\
    k_x \omega & \omega^2 - \mb^2
  \end{pmatrix}_{ab}
\end{equation}

\subsection{Photon propagating perpendicularly to the magnetic field}
\label{sec:propagating-photon}

Let us consider the case when a photon is propagating along the $z$ axis with
the magnetic field pointing in $x$ direction. We have:
\begin{equation}
  \label{eq:18}
  \Pi_{ab}  = -\frac{\k^2 H^2}{k^2 - \mb^2}\eta_{ab} -\frac{\k^2 H^2}{\mb^2(k^2 -
    \mb^2)}
  \begin{pmatrix}
    0  & 0\\
    0 & \omega^2
  \end{pmatrix}
\end{equation}
and one can find explicitly the propagator $G_{\mu\nu}$. A straightforward
computation shows that in this case
the $\{x,y\}$ block of the matrix $G_{\mu\nu}$
is diagonal and therefore the components
$A_x$ and $A_y$ are determined by $G_{xx}$
and $G_{yy}$ correspondingly. It is then easy to
see that the propagator $G_{xx}$ (describing wave with the electric field
along the magnetic field $H$) changes, while  $G_{yy}$
(polarization, orthogonal to the magnetic field) remains unchanged!  Namely,
$G_{xx}$ is given by
\begin{equation}
  \label{eq:19}
  G_{xx}(\omega,k_z) = -\frac{(k^2
    -\mb^2)}{(k^2-\m\gamma^2+i\epsilon)(k^2-\mb^2+i\epsilon)+\frac{\k^2
      H^2}{\mb^2}(\mb^2 -\omega^2)}\;,
\end{equation}
where $k^2 = \omega^2 - \kz^2$.

The zeros of the denominator are given by the following equation:
\begin{equation}
  \label{eq:20}
  (k^2-\m\gamma^2)(k^2-\mb^2)+\frac{H^2}{\vb^2}(\mb^2 -\omega^2)=0
\end{equation}
(where we have used $\mb = \k \vb$).  This equation describes two
propagating modes, one with a mass close to that of the photon:
\begin{equation}
  \label{eq:21}
  \omega_\gamma^2(k_z) \simeq k_z^2 (1-\eta) + \m\gamma^2 + \eta\mb^2\\
\end{equation}
and another with a mass, close to that of $B$ boson
\begin{equation}
  \label{eq:22}
  \omega_B^2(k_z) \simeq k_z^2 (1+\eta) + \mb^2
\end{equation}
Both solutions are found in the leading approximation in $(H/\vb^2)$ and
$(\m\gamma/\mb)$ and we have introduced a small parameter $\eta =
{H^2}/{\vb^4}$. Substituting these modes into the propagator~(\ref{eq:19}),
we see that the admixture of the ``heavy mode''~(\ref{eq:22}) is suppressed
by the small parameter $\eta$ and in the first approximation we obtain the
photon propagating with the mass
\begin{equation}
\label{eq:1111}
 m_{\gamma,H}^2  \simeq \frac{H^2}{\vb^2}
\end{equation}

Let us compare the result~(\ref{eq:19})--(\ref{eq:20}) with the similar
``secular equation'' of Ref.~\cite{Maiani:86}:
\begin{equation}
  \label{eq:23}
  k^2(k^2 - \m\phi^2) - \frac{H^2}{f_\phi^2}\omega^2=0
\end{equation}
We see that there is a trivial difference due to $\m\gamma$ and $(\omega^2 -
\mb^2)$ instead of $\omega^2$ in the last term of Eq.~(\ref{eq:20}).
However, we expect $\omega\gg \mb$ (in
experiments $\omega \sim 1\ev$). Therefore, the effects for a photon,
propagating in a perpendicular magnetic field,  remain in our model the same
as in the model with the axion if one identifies
\begin{equation}
  \label{eq:24}
 f_\phi \to \vb ;\qquad m_\phi \to \mb
\end{equation}
Using the PVLAS values for these constants we obtain:
\begin{equation}
  \label{eq:25}
  \mb \sim 10^{-3}\ev; \qquad \vb \sim 10^5\gev;\qquad \k \sim 10^{-17};\quad
  \v\gamma \sim \frac{\m\gamma}{\k} \lesssim 1\ev
\end{equation}
The exchange by this new light boson leads in principle to
corrections on the gravitational attraction of two bodies. The
mass and coupling constant (\ref{eq:25}) are compatible with the
existing constraints from the searches of such a ``fifth force" with a
range smaller than $0.1$ cm (in our case $\mb^{-1} \sim 0.01$ cm) (see e.g.
\cite{Fischbach:96}).

\section{Discussion}
\label{sec:discussion}

In this work we proposed a simple alternative to the axion model,
which reproduces the effects of rotation of the polarization plane,
recently observed by PVLAS collaboration, and avoids the
constraints coming from astrophysical and cosmological data. This
low-energy theory predicts the existence of a new light vector
field (rather than a pseudoscalar particle), which interacts with
the photon via Chern-Simons-like terms. Such terms generically
appear as effective interactions in theories with chiral
fermions due to fermionic loop
effects~\cite{DHoker:84a,DHoker:84b}. The structure of
interactions in Lagrangians similar to~(\ref{eq:7}) is
dictated by the requirement of anomaly cancellation and the
coefficients (\k in our case) are uniquely
fixed. For example, such terms would appear in the model with
fermions charged as shown in Table~\ref{tab:charges}.


 \begin{table}[t]
  \centering
  \begin{tabular}[c]{|>{$}c<{$}|>{$}c<{$}|>{$}c<{$}|>{$}c<{$}|>{$}c<{$}|>{$}c<{$}|>{$}c<{$}|>{$}c<{$}|>{$}c<{$}|>{$}c<{$}|>{$}c<{$}|>{$}c<{$}|>{$}c<{$}|}
    \hline
    & \phi_1 & \phi_2 & \psi_L & \psi_{R}& \psi_L' & \psi_{R}' &\chi_L &
    \chi_R & \chi_L' & \chi_R'\\
    \hline
    \u1_\gamma{} & \k{} & 0 & \q1 & \q1+\k{} & \q1+\k{} &
    \q1 & \q2 & \q2 & \q2' & \q2'\\
    \hline
    \u1_B & 0 & -\k{}& \beta_{1} & \beta_{1} & \beta_{1}' & \beta_{1}'
    &\beta_{2} & \beta_{2}-\k{}  & \beta_2 -\k{} & \beta_{2}
    \\\hline
  \end{tabular}
 \caption{A simple choice of charges of fermions and Higgs fields, which leads to the
   low-energy effective action~(\ref{eq:7}). Potential for Higgs fields
   ensure that $\phi_1 = \v\gamma e^{i\theta_1}$ and $\phi_2 = \vb e^{i\theta_2}$.
   The charges are chosen such that there are no $\gamma^3$
and $B^3$ triangle anomalies for both groups of fermions $\psi$
and $\chi$. On the other hand, mixed anomalies cancel
between the two groups of fermions.  Masses of $\psi$-fermions are generated
via Yukawa interaction with the Higgs field $\phi_1$ and masses of $\chi$-fermions by
the Higgs field $\phi_2$.}
  \label{tab:charges}
\end{table}

Clearly, for such a model to become realistic, one needs to implement it  in
an extension of the Standard Model (SM).  The realization of the model~(\ref{eq:7}) as
an effective theory derived from a consistent extension of the SM will be reported separately~\cite{Antoniadis:06}.  We would like to mention here that, apart from
integrating out four-dimensional fermions \`a la  D'Hoker and Fahri~\cite{DHoker:84a,DHoker:84b},
axionic and Chern-Simons terms can also appear from higher-dimensional theories.
For example, such effective
theories generically appear in realizations of SM on D-branes in string
theory~\cite{stringCS}.

In this work, we presented the simplest example of appearance of axion-like interactions as a
consequence of cancellation of \emph{gauge} anomalies between various sectors
of the theory.  Physically, this is a different approach to the building of such
models, as compared to the more conventional axions related to the strong CP
problem.  Other models of this kind can appear in various realistic scenarios,
which will also be discussed in~\cite{Antoniadis:06}.

There exists another class of theories, where effects, similar to those
discussed in this paper, may appear. In theories with extra dimensions
anomaly cancellation may occur between light particles, living in 4 dimensions
and very heavy particles, propagating in the higher-dimensional space (the so
called \emph{anomaly inflow} mechanism~\cite{Faddeev:85,Callan:84}). The
low-energy limit of such a theory may not have a description in terms of a
local 4-dimensional effective theory. The mechanism of anomaly inflow is
realized in nature in the case of quantum Hall
effect~\cite{Wen:90,Frohlich:91} where the 2-dimensional anomaly of the edge
excitations of the quantum Hall droplet is canceled by the inflow from a
3-dimensional Chern-Simons term. A 4-dimensional example of such a theory was
considered in~\cite{anomaly-th,anomaly-exp}. It was shown that the
anomaly cancellation between a 4-dimensional SM-like theory and a
5-dimensional bulk theory may lead to an effect similar to~(\ref{eq:5}), i.e.
appearance of a mass for the photon in the presence of strong magnetic field. The key
difference between  the effect described in~\cite{anomaly-th,anomaly-exp} and
the effects considered in this work is the dependence of the mass on the
magnetic field. Namely, this dependence had the form
\begin{equation}
  \label{eq:39}
  m_{\gamma,H} \sim \sqrt{\alpha_\EM\kappa_0 H}
\end{equation}
where $\alpha_\EM$ is the fine-structure constant and $\kappa_0$ is a
dimensionless constant, characterizing the deviation of the charges of light
fermions from their anomaly-free values (similar to \k of our model).
Numerically, this effect is consistent with PVLAS
data.  The different from~(\ref{eq:5}) dependence of the induced photon mass
on the magnetic field can thus serve as a peculiar signature of the presence
of extra dimensions. Thus, the measurement of the magnetic field dependence
of the effects of birefringence and dichroism can distinguish between various models of
``physics beyond the Standard Model''.

\begin{figure}[t]
  \centering
  \includegraphics[width=.5\textwidth]{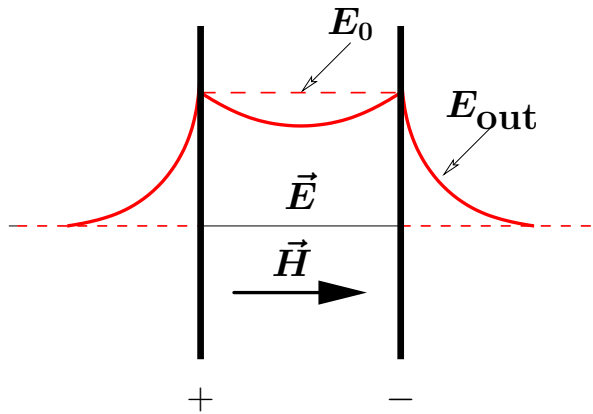} %
  \caption{Capacitor in a magnetic field. The presence of extra dimensions or
  axion makes electric field massive. As a result electric field in the capacitor
  decreases; however, one can detect the presence of a non-zero electric field outside
    the plates of the capacitor.}
  \label{fig:capacitor}
\end{figure}
To observe this effect, another type of experiment has been
suggested in~\cite{anomaly-exp}. Namely, by placing a capacitor in
a strong magnetic field, one can detect the redistribution of its
electric field, in particular the appearance of an electric field
outside the plates of a large capacitor
(Fig.~\ref{fig:capacitor}). In fact a similar effect would also be
present in theories with ALP (this is discussed briefly in
Appendix). We do not analyze here the feasibility of such an
experiment, however, we want to notice that unlike the effects of
rotation of polarization, the electric field $E_\mathrm{out} \sim
m_{\gamma,H}$ depends \emph{linearly} and not quadratically on the
induced mass of the photon and therefore could in principle be
stronger. Such an experiment would provide an independent test of
the interpretation of PVLAS data.  Moreover, this effect allows to
differentiate between various types of models, determine
separately mass and coupling constant of the axion, while its
dependence on $H$ can differentiate between the 4-dimensional
axion-like models and models with extra dimensions.


\subsection*{Acknowledgements}
We would like to thank L.~Alvarez-Gaum\'e, P.~Sikivie,
M.~Shaposhnikov, and I.~Tkachev for useful discussions.
This work was supported in part by the European Commission under the RTN
contract MRTN-CT-2004-503369 and in part by the INTAS contract 03-51-6346.

\appendix

\section{Capacitor effect for one axion }
\label{sec:capacitor}

\subsection{Massless  axion}
\label{sec:massless-axion}

Consider the Lagrangian of massless axion, interacting with the
electromagnetic field
\begin{equation}
  \label{eq:26}
  S_\mathrm{axion}=\int -\frac14 |F|^2 + \frac 12 |d\phi|^2 + \frac{\phi}{4f_\phi}
  F_{\mu\nu}\tilde F^{\mu\nu}
\end{equation}
In this case, equations of motion read
\begin{align}
  \label{eq:27}
  \p\nu F^{\mu\nu} & = j_0^\mu + \frac{1}{2f_\phi}\epsilon^{\mu\nu\lambda\rho}
  \p\nu
  \phi F_{\lambda\rho} \\
  \Box  \phi  & = \frac1{f_\phi} \epsilon^{\mu\nu\lambda\rho} F_{\mu\nu}
  F_{\lambda\rho}\label{eq:28}
\end{align}
Their solution can be easily found in the case when the source term $j_0$ is just
a (static) charge density $\rho(x)$ and the system is in the background
constant magnetic field $H^x$. We choose static gauge, so that only $A_0$ is
non-zero and solve Eq.~(\ref{eq:28}) by writing
\begin{equation}
  \label{eq:29}
  \p x  \phi  = \frac{H}{f_\phi} A_0
\end{equation}
which reduces Eq.~(\ref{eq:27}) to the Poisson equation
\begin{equation}
  \label{eq:30}
  \p x^2 A_0 = \rho(x) + \frac{H^2}{f_\phi^2} A_0
\end{equation}
with the mass of electric field being
\begin{equation}
  \m{\gamma,H}^2 = \frac{H^2}{f_\phi^2}.
\label{eq:31}
\end{equation}
We see that massless axion generates mass for static electric
field (in the presence of a strong magnetic field) and thus in a
capacitor experiment, similar to the one described in the section~\ref{sec:discussion}
(Fig. \ref{fig:capacitor}). The electric field \emph{outside} the
plates of the capacitor is given by $E_{out} \sim \m{\gamma,H}
V_0$, where $V_0$ is the voltage applied to the plates of the
capacitor. Unlike the effects of propagation of light in the
magnetic field, the effect here (nonzero $E_{out}$) is
proportional to the first power of magnetic mass $\m{\gamma,H}$.

\subsection{Massive axion}
\label{sec:massive-axion}

Let us now consider modifications compared to the previous
section~(\ref{sec:massless-axion}) in the
case when the axion has a mass $m_\phi$. Then, eqs.(\ref{eq:27})--(\ref{eq:28})
become:
\begin{align}
\label{eq:32} \p\nu F^{\mu\nu} & = j_0^\mu +
\frac{1}{2f_\phi}\epsilon^{\mu\nu\lambda\rho} \p\nu
 \phi  F_{\lambda\rho} \\
\bigl(\Box +m_ \phi ^2\bigr)  \phi  & = \frac1{4f_\phi}
\epsilon^{\mu\nu\lambda\rho} F_{\mu\nu}
F_{\lambda\rho}\label{eq:33}
\end{align}
Again, under the same assumptions (static charges $\rho(x)$, uniform background
magnetic field) we get the following solution in Fourier space ($-i\p x \to
k$)
\begin{equation}
  \label{eq:34}
   \phi (x) = \frac{ H}{f_\phi}\int \frac{dk}{2\pi}  \frac{i k e^{i k x}}{(k^2 + m_ \phi ^2)}A_0(k)
\end{equation}
and for $A_0$:
\begin{equation}
  \label{eq:35}
  A_0(x) = \int \frac{dk}{2\pi} \frac{e^{i k x} k^2}{(k^2 + m_ \phi ^2)(k^2 +
    \m{\gamma,H}^2)} \rho(k)
\end{equation}
This integral is convergent and in the limit $m_ \phi \to 0$ one
recovers the solution of Eq.~(\ref{eq:30}) of the massless axion.
Let us analyze it for the simplest case $\rho(x) = q_0 \delta(x) $
or $\rho(k) = q_0$:
\begin{equation}
  \label{eq:36}
  A_0(x) =\frac{q_0}{4\pi}\left[e^{-m_\phi |x|}\frac{m_\phi}{m_\phi^2 - \m{\gamma,H}^2} +
    e^{-\m{\gamma,H} |x|}\frac{m_\phi}{\m{\gamma,H}^2 - m_\phi^2}\right]
\end{equation}
In the interesting case $\m{\gamma,H}  \ll m_ \phi $ one finds
\begin{equation}
  \label{eq:37}
  A_0(x) \approx A_\mathrm{massless}(x)\left\{
    \begin{aligned}
      \frac{\m{\gamma,H}^2}{m_\phi^2} , && |x|&\gg \frac1{m_\phi}\\
      \frac{\m{\gamma,H}}{m_\phi} , && |x| &\ll \frac1{m_\phi}
    \end{aligned}
    \right.
\end{equation}
i.e. the expression for the case of zero axion mass, suppressed by
some power of $\m{\gamma,H}/m_\phi$.

\providecommand{\href}[2]{#2}\begingroup\raggedright\endgroup

\end{document}